\documentclass[12pt,sec]{article}
\usepackage{epsfig}
\usepackage{axodraw}
\begin{document}

\begin{center}
{\large  Understanding the $e^+e^-\rightarrow D^{(*)+}D^{(*)-}$ processes observed by Belle }\\[0.8cm]
{ Kui-Yong Liu}

{\footnotesize Department of Physics, Peking University,
 Beijing 100871, People's Republic of China
 and Department of Physics, Liaoning University, Shenyang 110036, People's Republic of China}\\[0.5cm]

{ Zhi-Guo He and Yu-Jie Zhang}

{\footnotesize Department of Physics, Peking University,
 Beijing 100871, People's Republic of China}\\[0.5cm]
{ Kuang-Ta Chao}

{\footnotesize China Center of Advanced Science and Technology
(World Laboratory), Beijing 100080, People's Republic of China and
Department of Physics, Peking University,
 Beijing 100871, People's Republic of China}

\end{center}

\begin{abstract}

We calculate the production cross sections for $D^{*+}D^{*-}$,
$D^+D^{*-}$ and $D^+D^-$ in $e^+e^-$ annihilation through one
virtual photon in the framework of perturbative QCD with
constituent quarks. The calculated cross sections for
$D^{*+}D^{*-}$ and $D^+D^{*-}$ production are roughly in agreement
with the recent Belle data. The helicity decomposition for $D^{*}$
meson production is also calculated. The fraction of the
$D^{*\pm}_LD^{*\mp}_T$ final state in $e^+e^-\rightarrow
D^{*+}D^{*-}$ process is found to be 65\%. The fraction of
$DD^*_T$ production is 100\% and $DD^*_L$ is forbidden in $e^+e^-$
annihilation through one virtual photon. We further consider
$e^+e^-$ annihilation through two virtual photons, and then find
the fraction of $DD^{*}_T$ in $e^+e^-\rightarrow DD^{*}$ process
to be about 91\%.

PACS number(s): 12.40.Nn, 13.85.Ni, 14.40.Gx

\end{abstract}

\begin{figure}[t]
\begin{center}
\vspace{-3.0cm}
\includegraphics[width=12cm,height=16cm]{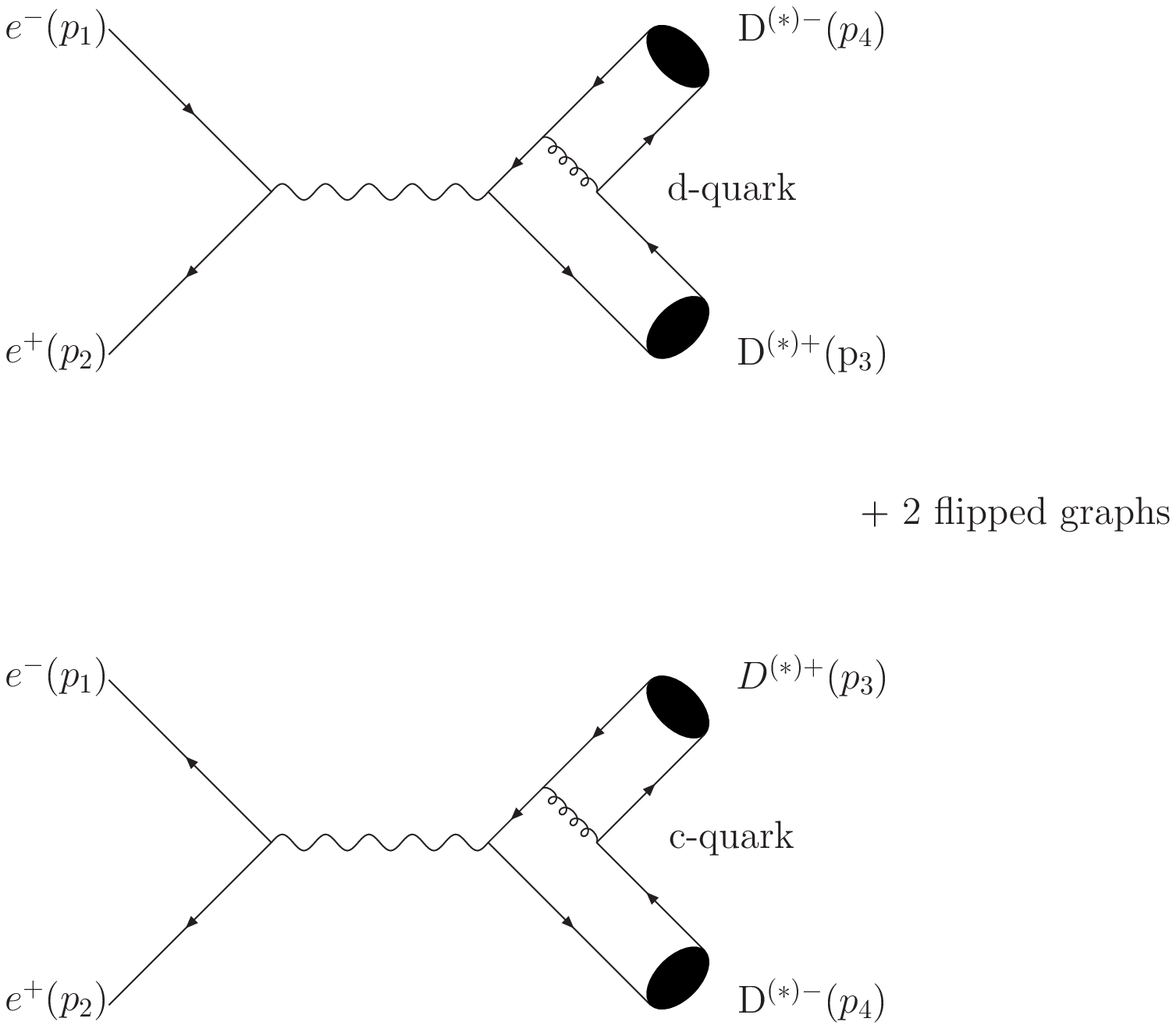}
\vspace{-3cm}
\end{center}
\caption{ Feynman diagrams for $e^+e^-\rightarrow \gamma^*
\rightarrow D^{(*)+}D^{(*)-}$.} \label{fey}
\end{figure}

Recently, heavy meson production in $e^+e^-$ annihilation has
become a very interesting subject both experimentally and
theoretically. For instance, for charmonium production in $e^+e^-$
annihilation at the B-factory energy $\sqrt{s}=10.6$GeV, there are
large differences between the experimental
data~\cite{babar,belle,exdou} and the calculated cross sections
for both exclusive processes~\cite{ex} and inclusive
processes~\cite{in}. Even by including the effects of $e^+e^-$
annihilation into two photons in the exclusive double-charmonium
production~\cite{double} and inclusive charmonium
production~\cite{two}, the large discrepancies still
exist~\cite{comment}. Moreover, most recently the Belle
Collaboration has measured \cite{uglov} the charmed meson pair
$D^{(*)+}D^{(*)-}$ production in $e^+e^-$ annihilation and also
found a large differences between observed data and theoretical
predictions\cite{grozin}. The measured cross sections are
$\sigma(e^+e^-\rightarrow D^{{*}+}D^{{*}-})=0.65\pm0.04\pm0.07~$pb
and $\sigma(e^+e^-\rightarrow
D^{+}D^{{*}-})=0.71\pm0.05\pm0.09~$pb, which are, however, lower
than those predicted in Ref.\cite{grozin} by an order of magnitude
(see \cite{uglov} for the comparison). The Belle measurements of
the exclusive D meson pair production in $e^+e^-$ annihilation
could be another challenge to the theoretical studies. In order to
understand these production processes, in this letter we will
present a calculation in the framework of perturbative QCD with
constituent quarks. Namely, we will treat the charmed mesons
$D^{(*)}$ as bound states of a constituent charmed quark and a
constituent light antiquark, and the virtual photon will couple to
the charm quark, which will subsequently emit a light quark pair
with constituent quark masses through a virtual gluon (see the
upper diagram in Fig.~\ref{fey}). The virtuality of the virtual
gluon could be large enough for the application of perturbative
QCD. In addition, the case of the virtual photon coupled to the
light quark will also be considered (see the lower diagram in
Fig.~\ref{fey}). In the following we will report our calculated
results for both the total cross sections and the helicity
decomposed cross sections for these charmed meson pair production
in $e^+e^-$ annihilation at $\sqrt{s}=10.6$GeV, .

Following the method in Ref.~\cite{amp}, the amplitude for
producing the heavy quark $(Q)$ -light antiquark $(\bar{q})$
bound-state $(Q\bar{q})$ is given by

\begin{equation}
\label{amp} A(P)=\sum_{L_Z, S_Z}\int\frac{d^3{\bf
k}}{(2\pi)^3}\Psi_{L,L_Z}({\bf k})\langle LL_Z;SS_Z\mid
JJ_Z\rangle M(P,k),
\end{equation}
where
\begin{equation}
M(P,k)={\cal O}_\Gamma~ \Gamma_{SS_Z}(P,k),
\end{equation}
and ${\cal O}_{\Gamma}$ represents the short-distance interaction
producing the Q and $\bar{q}$ in a specific bound-state. {\bf k}
is the relative momentum between Q and $\bar{q}$.
$\Psi_{L,L_Z}({\bf k})$ is the Bethe-Salpeter bound state
wave-function. $\Gamma_{SS_Z}(P,k)$, up to second order in k, is
given by

\begin{eqnarray}
&\Gamma_{SS_Z}(P,k)&=\sqrt{\frac{m_Q+m_q}{2m_Qm_q}}\sum_{s,\bar{s}}\langle
\frac{1}{2}s;\frac{1}{2}\bar{s} \mid SS_Z\rangle v(rP-k,s)\bar{u}(\bar{r}P+k,\bar{s}),\nonumber\\
&&\approx \sqrt{m_Q+m_q}\left (\frac {r\not{P}-\not{k}-m_q}{2m_q}
\right ) \left (\begin{array}{c}
  \gamma^5 \\
  -\not{\epsilon}(P,S_Z) \\
\end{array}  \right ) \left
(\frac{\bar{r}\not{p}+\not{k}+m_Q}{2m_Q} \right )
\end{eqnarray}
where $\gamma^5$ is for spin-singlet state and
$-\not{\epsilon}(P,S_Z)$ is for spin-triplet state.

For the S-wave state, we have $M(P,k)=M(P,0)$. After integrating
over {\bf k}, the amplitude $A(P)$ is related to the origin of the
radial wave-function $R_S(0)$

\begin{equation}
\label{origin} \int \frac{d^3 {\bf
k}}{(2\pi)^3}\Psi_{00}=\frac{R_S(0)}{\sqrt{4\pi}}.
\end{equation}

Using Eq.~(\ref{amp}-\ref{origin}), the calculation of the cross
section for the process showed in Fig.~\ref{fey} is
straightforward. We get the cross section as follows

\begin{equation}
d\sigma=\frac{\pi
\alpha^2\alpha_s^2R_s(0)^4}{8s}\sqrt{1-\frac{4m^2}{s}}\mid \bar{M}
\mid ^2dx,
\end{equation}
where $x=\cos\theta$, $\theta$ is the angle between ${\bf p_1}$
and ${\bf p_3}$. The meson mass m equals $m_c+m_d$ in the leading
oeder non-relativistic approximation. For the $e^+e^-\rightarrow
D^{*+}D^{*-}$ process, $\mid \bar{M} \mid ^2$ reads

\begin{eqnarray}
&\mid \bar{M} \mid^2 &=\frac{128(m_c+m_d)^2(4m_c^2+8m_cm_d+4m_d^2-s)}{81m_c^6m_d^6s^5}\nonumber\\
&&\times[-48m_c^{12}-192m_c^{11}m_d-288m_c^{10}m_d^2-192m_c^9m_d^3-96m_c^8m_d^4-192m_c^7m_d^5\nonumber\\
&&-288m_c^6m_d^6-192m_c^5m_d^7-60m_c^4m_d^8-48m_c^3m_d^9-72m_c^2m_d^{10}-48m_cm_d^{11}\nonumber\\
&&-12m_d^{12}-8m_c^{10}s-48m_c^9m_ds-80m_c^8m_d^2s-64m_c^7m_d^3s-56m_c^6m_d^4s\nonumber\\
&&-64m_c^5m_d^5s-50m_c^4m_d^6s-28m_c^3m_d^7s-20m_c^2m_d^8s-12m_cm_d^9s-2m_d^{10}s\nonumber\\
&&-4m_c^6m_d^2s^2-4m_c^4m_d^4s^2-m_c^2m_d^6s^2+(48m_c^{12}+192m_c^{11}m_d+288m_c^{10}m_d^2\nonumber\\
&&+192m_c^{9}m_d^3+96m_c^{8}m_d^4+192m_c^{7}m_d^5+288m_c^{6}m_d^6+192m_c^{5}m_d^7+60m_c^{4}m_d^8\nonumber\\
&&+48m_c^{3}m_d^9+72m_c^{2}m_d^{10}+48m_cm_d^{11}+12m_d^{12}-8m_c^{10}s-16m_c^{9}m_ds-16m_c^{8}m_d^{2}s\nonumber\\
&&-16m_c^{7}m_d^{3}s-24m_c^{6}m_d^{4}s-32m_c^{5}m_d^{5}s-18m_c^{4}m_d^{6}s-4m_c^{3}m_d^{7}s\nonumber\\
&&-4m_c^{2}m_d^{8}s-4m_cm_d^{9}s-2m_d^{10}s+4m_c^6m_d^2s^2+4m_c^4m_d^4s^2+m_c^2m_d^6s^2)x^2].
\end{eqnarray}

For the $e^+e^-\rightarrow D^{+}D^{*-}$ process, one should
replace $\sqrt{1-\frac{4m^2}{s}}$ with
$\sqrt{\frac{(s+m_{D^+}^2-m_{D^{(*)-}}^2)^2}{s^2}-\frac{4m^2}{s}}$
in the phase space integration. $\mid \bar{M} \mid ^2$ is given by
\begin{eqnarray}
\mid \bar{M} \mid^2
&=&-\frac{128(m_c+m_d)^6(2m_c^3-m_d^3)^2(4m_c^2+8m_cm_d+4m_d^2-s)(1+x^2)}{81m_c^6m_d^6s^4}.
\end{eqnarray}

For the $e^+e^-\rightarrow D^{+}D^{-}$ process, $\mid \bar{M} \mid
^2$ is as follows

\begin{eqnarray}
&\mid \bar{M} \mid^2&
=-\frac{128(m_c+m_d)^2(4m_c^2+8m_cm_d+4m_d^2-s)(1-x^2)}{81m_c^6m_d^6s^4}\nonumber\\
&&\times(4m_c^6+8m_c^5m_d+4m_c^4m_d^2+2m_c^2m_d^4+4m_cm_d^5+2m_d^6-2m_c^3m_ds-m_cm_d^3s).
\end{eqnarray}

The Mandelstam variables are defined as
\begin{equation}
\label{t} t=(p_3-p_1)^2=m^2-\frac{s}{2}\left
(1-\sqrt{1-\frac{4m^2}{s}}x \right ),
\end{equation}
\begin{equation}
\label{u} u=(p_3-p_2)^2=m^2-\frac{s}{2}\left
(1+\sqrt{1+\frac{4m^2}{s}}x \right ).
\end{equation}

We notify that for $e^+e^-\rightarrow D^{+}D^{*-}$ process, the
formula of Mandelstam variables t and u are a little different
from Eq.~(\ref{t}) and Eq.~(\ref{u}) because of the mass
difference between $D^{+}$ and $D^{*-}$. As the difference is
small, the change for the cross section is negligible. For
simplicity of the module of amplitude, we still use the definition
in Eq.~(\ref{t}) and Eq.~(\ref{u}).

In the numerical calculation, the input parameters are as follows

\begin{equation}
\alpha=1/137,~~~~~~\alpha_s=0.26.
\end{equation}

For the Coulomb-plus-linear potential case, the value of the wave
function at the origin for charmonium can be found e.g. in
\cite{quigg}. With $m_c=1.84$ GeV, for the S-wave charmonium it
could be given by

\begin{equation}
|R_S(0)|^2=1.454~{\rm GeV^3}.
\end{equation}

We then use the potential scaling rules to get a rough estimate
for wave function at the origin for the charmed meson. We fix the
constituent c-quark mass at 1.6 GeV. The cross sections for
$D^{(*)+}D^{(*)-}$, $D^{+}D^{(*)-}$ and $D^{+}D^{-}$ are found to
be

\begin{equation}
\label{sgdsds}\sigma(e^+e^-\rightarrow \gamma^* \rightarrow
D^{{*}+}D^{{*}-})=0.532~{\rm pb},
\end{equation}

\begin{equation}
\sigma(e^+e^-\rightarrow \gamma^* \rightarrow
D^{+}D^{{*}-})=0.699~{\rm pb},
\end{equation}

\begin{equation}
\sigma(e^+e^-\rightarrow \gamma^* \rightarrow
D^{+}D^{-})=0.098~{\rm pb}.
\end{equation}

We also calculate the polarized $D^{(*)+}D^{(*)-}$ and
$D^{+}D^{(*)-}$ production. The polarized cross section can be
calculated by defining the longitudinal polarization vector as
follows~\cite{pol}

\begin{equation}
\epsilon^{\mu}_L(p)=\frac{p^{\mu}}{M}-\frac{Mn^{\mu}}{n\cdot p},
\end{equation}
where $p^2=M^2$ and $n^{\mu}=(1,-\vec{p}/\mid \vec{p} \mid)$.

The polarized modules of amplitude for $e^+e^-\rightarrow \gamma^*
\rightarrow D^{(*)+}D^{(*)-}$ is listed in Eq.~\ref{onephoton}
\begin{eqnarray}
\label{onephoton}\mid \bar{M}
\mid^2=\frac{-128(m_c+m_d)^6(2m_c^3+m_d^3)^2(4m_c^2+8m_cm_d+4m_d^2-s)(1+x^2)}{81m_c^6m_d^6s^4},
\end{eqnarray}

The corresponding cross sections are

\begin{equation}
\label{pol1}\sigma(e^+e^-\rightarrow \gamma^* \rightarrow
D^{(*)\pm}_LD^{(*)\mp}_T)=0.347~{\rm pb},
\end{equation}

Using the results in Eq.~(\ref{sgdsds}) and~(\ref{pol1}), we get
\begin{equation}
\frac{\sigma(e^+e^-\rightarrow
D^{(*)\pm}_LD^{(*)\mp}_T)}{\sigma(e^+e^-\rightarrow
D^{(*)+}D^{(*)-})}=65.2\%
\end{equation}

For $D^+D^{(*)-}$ case, the longitudinal cross section for one
photon process must be zero, since it is forbidden by parity
conservation and angular momentum conservation. The effective
$\gamma DD^{*}$ vertex in Fig~\ref{fey} only can have the form as

\begin{equation}
{\cal
L}_{int}\sim\varepsilon^{\mu\nu\rho\sigma}F_{\mu\nu}(q)G_{\rho\sigma}(p_4)\phi(p_3),
\end{equation}
where $F_{\mu\nu}(q)$ represents the field of the virtual photon,
$G_{\rho\sigma}(p_4)$ represents the field of the $D^{(*)-}$ meson
and $\phi(p_3)$ represents the field of the $D^+$ meson. In the
momentum space, the effective vertex is as follow

\begin{equation}
\label{int}{\cal
L}_{int}\sim\varepsilon^{\mu\nu\rho\sigma}q_{\mu}\epsilon^{\lambda}_{\nu}(q)
p_{4\rho}\epsilon^{\lambda^{'}}_{\sigma}(p_4).
\end{equation}

 If we choose the center-of-mass
frame of the $e^+e^-$, the formula in Eq.(\ref{int}) becomes

\begin{eqnarray}
\label{ppp}{\cal
L}_{int}&\sim&\varepsilon^{0\nu\rho\sigma}q_{0}\epsilon^{\lambda}_{\nu}(q)
p_{4\rho}\epsilon^{\lambda^{'}}_{\sigma}(p_4)\nonumber\\
&\sim& q_{0}(\vec{p}_4
\times\vec{\epsilon}^{\lambda^{'}}(p_4))\cdot\vec{\epsilon}^{\lambda}(q).
\end{eqnarray}

\begin{figure}[t]
\begin{center}
\vspace{-3.0cm}
\includegraphics[width=12cm,height=16cm]{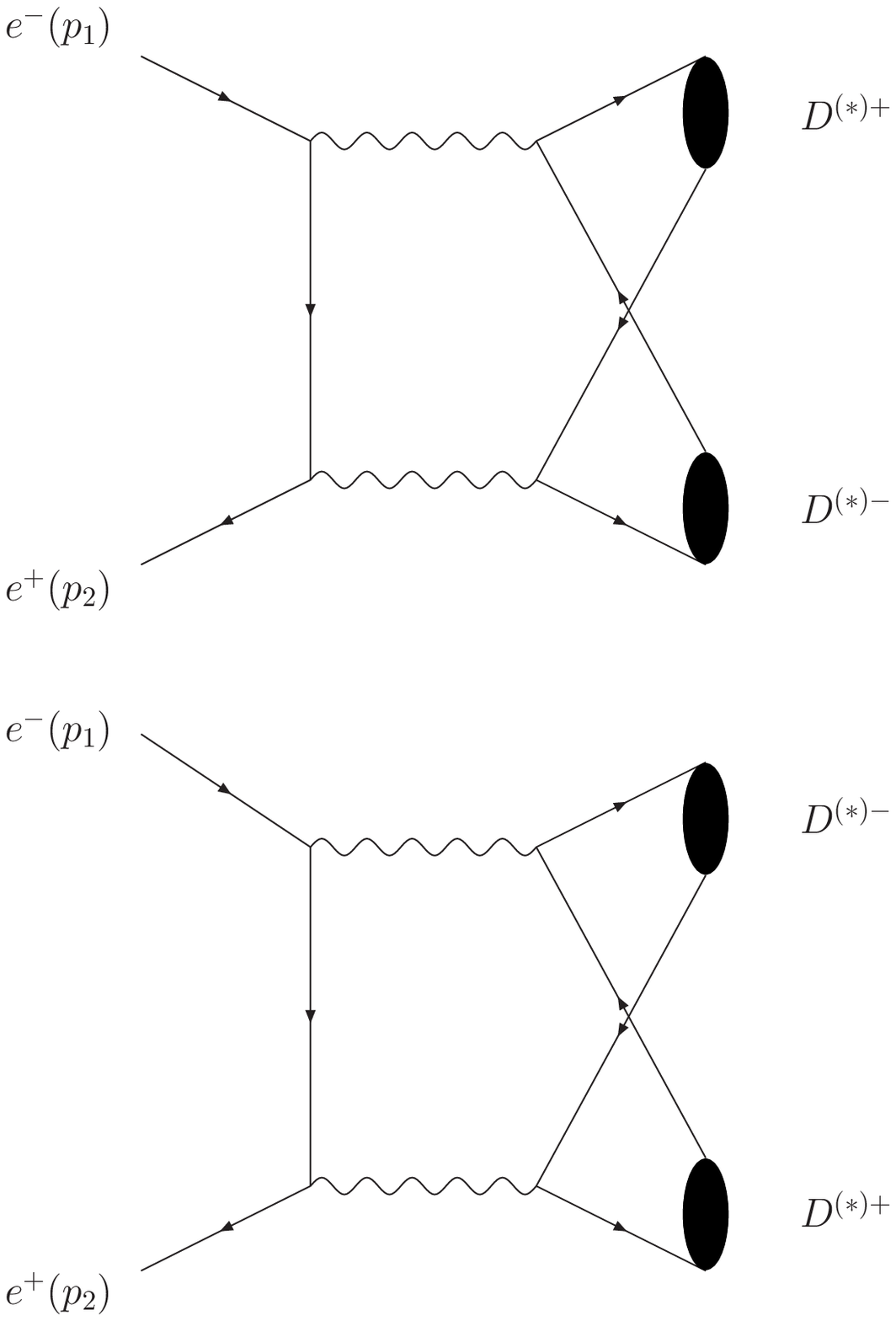}
\vspace{-3cm}
\end{center}
\caption{ Feynman diagrams for $e^+e^-\rightarrow 2\gamma^*
\rightarrow D^{(*)+}D^{(*)-}$.} \label{twofey}
\end{figure}

Because the space component of the longitudinal polarization
vector of the $D^*$ meson is parallel to the momentum of the $D^*$
meson, from Eq.(\ref{ppp}), one knows that the longitudinally
polarized $D^*$ meson cannot contribute to the process listed in
Fig.~\ref{fey}. The longitudinal $D^{*}$ only comes from two
photons process, as listed in Fig.~\ref{twofey}.

The modules of amplitude for $e^+e^-\rightarrow 2\gamma^*
\rightarrow D^+D^{(*)-}$ and $e^+e^-\rightarrow 2\gamma^*
\rightarrow D^+D^{(*)-}_T$ are listed, respectively, in
Eq.~(\ref{twodds}) and Eq.~(\ref{twopol}).

\begin{eqnarray}
&\label{twodds} \mid \bar{M} \mid & =
\frac{8(m_c-m_d)^2(m_c+m_d)^8(m_c^2+3m_cm_d+m_d^2)^2(4(m_c+m_d)^2(1-x^2)+s(1-x^2))}
{81m_c^6m_d^6(2m_c+m_d)^2(m_c+2m_d)^2s^3}\nonumber\\
\end{eqnarray}

\begin{equation}
\label{twopol}\mid
\bar{M}_T\mid^2=\frac{32(m_c-m_d)^{2}(m_c+m_d)^{10}(mc_2+3m_cm_d+m_d^2)^2(1+x^2)}
{81m_c^6m_d^6s^3(2m_c+m_d)^2(m_c+2m_d)^2}.
\end{equation}

The corresponding cross sections are

\begin{equation}
\label{cdds}\sigma(e^+e^-\rightarrow 2 \gamma^* \rightarrow
D^{+}D^{(*)-})=0.091~{\rm pb}
\end{equation}

\begin{equation}
\label{tdds}\sigma(e^+e^-\rightarrow 2\gamma^* \rightarrow
D^{+}D^{(*)-}_T)=0.018~{\rm pb}.
\end{equation}

Then one can get the ratio of cross section for $D_T^{*}$ to total
cross section as follows
\begin{equation}
\frac{\sigma(e^+e^-\rightarrow
D^{+}D^{(*)-}_T)}{\sigma(e^+e^-\rightarrow D^{+}D^{(*)-})}=90.8\%.
\end{equation}

In summary, motivated by the measurement of the $e^+e^-
\rightarrow D^{(*)+}D^{(*)-}$ cross-sections~\cite{uglov}, we
calculate the cross-sections for $D^{(*)+}D^{(*)-}$,
$D^{+}D^{(*)-}$ and $D^{+}D^{-}$ in $e^+e^-$ annihilation through
one photon. The calculated cross-sections are roughly in agreement
with the experimental data. The fraction of the
$D_L^{(*)+}D_T^{(*)-}$ final state in the $e^+e^-\rightarrow
D^{(*)+}D^{(*)-}$ reaction is also calculated, the ratio of 65\%
is in some deviation from the Belle data, which is $(97 \pm
5)\%$~\cite{uglov}. Moreover, for the $D^{+}D^{(*)-}$ production
case, in Ref.~\cite{uglov}, it is claimed that $e^+e^- \rightarrow
D^{+}D^{(*)-}$ is saturated by $D^{*}_L$ final state ($95.8 \pm
5.6)\%$, in good agreement with the predictions of
Ref.~\cite{grozin}. Under our analysis, the $D_L^{*}$ final state
is forbidden by parity conservation and angular momentum
conservation. If the $e^+e^-\rightarrow 2\gamma^* \rightarrow
D^{+}D^{(*)-}$ process is further considered, the fraction of the
$D^*_T$ final state in $e^+e^-\rightarrow D^{+}D^{(*)-}$ process
is 90.8\%.

\section*{Acknowledgments}

The authors thank J.P.Ma, C.Meng and Z.Z.Song for useful
discussions. This work was supported in part by the National
Natural Science Foundation of China, and the Education Ministry of
China.

\newpage

\newpage

\end{document}